\documentstyle[epsf]{mn}

\newif\ifAMStwofonts
\def\bout{B_{\rm out}}
\def\ceff{C_{\rm eff}}

\ifoldfss
  \ifCUPmtlplainloaded \else
    \NewTextAlphabet{textbfit} {cmbxti10} {}
    \NewTextAlphabet{textbfss} {cmssbx10} {}
    \NewMathAlphabet{mathbfit} {cmbxti10} {} % for math mode
    \NewMathAlphabet{mathbfss} {cmssbx10} {} %  "   "    "
  \fi
  \ifAMStwofonts
    \ifCUPmtlplainloaded \else
      \NewSymbolFont{upmath} {eurm10}
      \NewSymbolFont{AMSa} {msam10}
      \NewMathSymbol{\upi}     {0}{upmath}{19}
      \NewMathSymbol{\umu}     {0}{upmath}{16}
      \NewMathSymbol{\upartial}{0}{upmath}{40}
      \NewMathSymbol{\leqslant}{3}{AMSa}{36}
      \NewMathSymbol{\geqslant}{3}{AMSa}{3E}

    \fi
  \fi
\fi % End of OFSS

\ifnfssone
  \newmathalphabet{\mathit}
  \addtoversion{normal}{\mathit}{cmr}{m}{it}
  \addtoversion{bold}{\mathit}{cmr}{bx}{it}
  \newmathalphabet{\mathbfit} % math mode version of \textbfit{..}
  \addtoversion{normal}{\mathbfit}{cmr}{bx}{it}
  \addtoversion{bold}{\mathbfit}{cmr}{bx}{it}
  \newmathalphabet{\mathbfss} % math mode version of \textbfss{..}
  \addtoversion{normal}{\mathbfss}{cmss}{bx}{n}
  \addtoversion{bold}{\mathbfss}{cmss}{bx}{n}
  \ifAMStwofonts
    \ifCUPmtlplainloaded \else
      \UseAMStwoboldmath
      \makeatletter
      \new@mathgroup\upmath@group
      \define@mathgroup\mv@normal\upmath@group{eur}{m}{n}
      \define@mathgroup\mv@bold\upmath@group{eur}{b}{n}
      \edef\UPM{\hexnumber\upmath@group}
      \new@mathgroup\amsa@group
      \define@mathgroup\mv@normal\amsa@group{msa}{m}{n}
      \define@mathgroup\mv@bold\amsa@group{msa}{m}{n}
      \edef\AMSa{\hexnumber\amsa@group}
      \makeatother
      \mathchardef\upi="0\UPM19
      \mathchardef\umu="0\UPM16
      \mathchardef\upartial="0\UPM40
      \mathchardef\leqslant="3\AMSa36
      \mathchardef\geqslant="3\AMSa3E
    \fi
  \fi
\fi % End of NFSS release 1

\ifnfsstwo
  \DeclareMathAlphabet{\mathbfit}{OT1}{cmr}{bx}{it}
  \SetMathAlphabet\mathbfit{bold}{OT1}{cmr}{bx}{it}
  \DeclareMathAlphabet{\mathbfss}{OT1}{cmss}{bx}{n}
  \SetMathAlphabet\mathbfss{bold}{OT1}{cmss}{bx}{n}
  \ifAMStwofonts
    \ifCUPmtlplainloaded \else
      \DeclareSymbolFont{UPM}{U}{eur}{m}{n}
      \SetSymbolFont{UPM}{bold}{U}{eur}{b}{n}
      \DeclareSymbolFont{AMSa}{U}{msa}{m}{n}
      \DeclareMathSymbol{\upi}{0}{UPM}{"19}
      \DeclareMathSymbol{\umu}{0}{UPM}{"16}
      \DeclareMathSymbol{\upartial}{0}{UPM}{"40}
      \DeclareMathSymbol{\leqslant}{3}{AMSa}{"36}
      \DeclareMathSymbol{\geqslant}{3}{AMSa}{"3E}
    \fi
  \fi
\fi % End of NFSS release 2

\ifCUPmtlplainloaded \else
  \ifAMStwofonts \else % If no AMS fonts
    \def\upi{\pi}
    \def\umu{\mu}
    \def\upartial{\partial}
  \fi
\fi

\title[Early-type galaxies: Outflows versus SF Efficiency]
{On breaking the age-metallicity degeneracy in early-type galaxies: 
	Outflows versus Star Formation Efficiency}
\author[I. Ferreras \& J. Silk]
{Ignacio Ferreras \& Joseph Silk\thanks{\tt ferreras,silk@astro.ox.ac.uk}\\
Nuclear \& Astrophysics Lab. 1 Keble Road, Oxford OX1 3RH, 
	United Kingdom}

\date{Draft version \today}

\pagerange{\pageref{firstpage}--\pageref{lastpage}}
\pubyear{2000}

\begin{document}

\maketitle

\label{firstpage}

\begin{abstract}
A simple model of chemical enrichment in cluster early-type galaxies
is presented where the main parameters driving the formation of the
stellar component are reduced to four: infall timescale ($\tau_f$),
formation epoch ($z_F$), star formation efficiency ($\ceff$) and 
fraction of gas ejected in outflows ($\bout$). We find that only
variations in $\bout$ or $\ceff$ can account for the colour-magnitude 
relation, so that the most luminous galaxies had low values of ejected 
gas and high efficiencies. Less massive galaxies can be either 
related to a lower star formation efficiency ($\ceff$-sequence) or to
an increased outflow rate ($\bout$-sequence). The combination of
chemical enrichment tracks with population synthesis
models (Bruzual \& Charlot 2000) is used to explore the correlation
between mass-to-light ratios and masses. A significant slope mismatch is 
found between stellar and total $M/L$ ratios, which 
cannot be explained by an age spread and implies a
non-linear correlation between total and stellar mass:
$M_{\rm TOT} \propto M_{ST}^{1.2}$. The sequences driven by
star formation efficiency ($\ceff$) and outflows ($\bout$)
are shown to predict different trends at high redshift.
The variation with redshift of the slope of the fundamental 
plane will increase significantly in the efficiency sequence --- driven
by age --- and will slightly decrease in the outflow sequence --- driven
by metallicity. The evolution of the zero point is similar in 
both cases and within the observational errors of current 
observations.  Measurement of the dependence of the
tilt of the fundamental plane on redshift will break the degeneracy
between outflows and star formation efficiency, which will
enable us to determine whether the colour-magnitude relation is
controlled by age or metallicity. 
\end{abstract}


%%%%%%%%%%%%%%%%%%%%%%%%%%%%%%%%%%%%%%%%%%%%%%%%%%%%%%%%%%%%%%%%%%%%%%

\section{Introduction}

The process of galaxy formation and evolution can be explored
in two complementary ways: A ``forwards'' approach takes into
account the physics underlying the most basic processes of
structure evolution and star formation and --- after finding a
suitable set of initial conditions --- evolves the system
forward so that the final output is compared with observations.
This is the philosophy behind N-body simulations or
semi-analytic modelling (e.g. Baugh et al. 1998;
Kauffmann \& Charlot 1998). On the other hand, a 
``backwards''  approach simplifies the physics behind galaxy 
formation and evolution to a phenomenological problem comprising 
a reduced set of parameters, using local observations as constraints. 
Then the system is evolved backwards so that the predictions 
for a given set of parameters are compared with observations 
at moderate-to-high redshifts (e.g. Bouwens, Broadhurst \& Silk
1998a,b). This ``brute force'' method allows 
one to search a reasonable volume of parameter space, throwing 
light on processes, such as star formation, which are 
otherwise too complicated to tackle from basic principles.

Cluster early-type galaxies are ideal candidates for a
comparison between models and observations. Significant
samples of these galaxies can be found over a large
redshift range. Furthermore, the tight observed correlations
such as the colour-magnitude relation or the fundamental plane
can be used as powerful constraints on a phenomenological
backwards approach. The list of observed clusters at moderate
to high redshift is quite large (e.g. Dressler et al. 1999
; Stanford, Eisenhardt \& Dickinson 1998; Oke, Postman \& Lubin 1998;
Van Dokkum 1999) and ever increasing 
(e.g. Yee et al. 1999). Clusters
observed at redshifts $z\sim 1-1.5$ yield valuable information
about the epoch of star formation, pushing it to very high
redshift ($z_F\ga 3$). 

Unfortunately, the direct spectrophotometric determination of the 
star formation history is hampered by the age-metallicity degeneracy 
(Worthey 1994; Ferreras, Charlot \& Silk 1999) which allows the 
variations of most of the spectrophotometric 
observables to be explained either by a range of ages or metallicities. 
Broadband photometry is strongly dependent both on age and metallicity
but even spectral indices targeting single lines such as Balmer 
absorption or magnesium abundance can change both with age and metallicity.
Not surprisingly, a similar age estimation technique based on 
spectral indices applied to similar sets of elliptical galaxies
yields contradictory results: Kuntschner (2000) and Kuntscher \& Davies (1998) 
find coeval stellar populations in Fornax cluster ellipticals so that
the colour range is explained by a metallicity sequence. On the other 
hand, the sample of field and group ellipticals observed by Gonz\'alez (1993) 
and further analysed by Trager et al. (2000) presents 
a relatively large spread in ages. Hence, the issue of the stellar age 
distribution in galaxies still requires the aid of modelling. We will
show that incorporating chemical enrichment allows one to potentially
solve the age-metallicity degeneracy problem. The next
three sections describe our chemical enrichment model and the meaning
of the reduced set of parameters used to trace the star formation history
in cluster ellipticals. \S5 deals with the comparison of predicted
and observed mass-to-light ratios and its use at high redshift to
discriminate between a mass sequence driven by age or metallicity.
Finally in \S6 we discuss the predictions and list the main conclusions.

%%%%%%%%%%%%%%%%%%%%%%%%%%%%%%%%%%%%%%%%%%%%%%%%%%%%%%%%%%%%%%%%%%%%%

\section{Model description}

The basic mechanisms describing chemical enrichment in galaxies can
be reduced to infall of gas, metal-rich outflows triggered by 
supernovae, and a star formation prescription. 
Ferreras \& Silk (2000, hereafter FS00) described 
one such simple model for an exponentially decaying infall rate 
of primordial gas in terms of five parameters: flow rate, 
timescale and delay of infall, ejected fraction in outflows, and star 
formation efficiency. Monotonically decreasing star formation
rates (SFRs) always encounter the so-called G-dwarf 
problem. In FS00 the problem was avoided by assuming the 
metallicity effect was predominant with regard to the 
spectrophotometric output, thereby enabling us to convolve 
simple stellar populations with different metallicities given 
by the chemical enrichment equations but with an average common age.

The G-dwarf problem arises when comparing the predictions of closed box
models with the observed paucity of low-metallicity low-mass stars 
in the halo of our galaxy (Van den Bergh 1962). However,
this problem extends to all galaxies and morphologies 
(Worthey et al. 1996) since any model assuming 
a monotonically decreasing infall rate overproduces 
low-metallicity, low-mass stars. These models can be readily ruled
out for bright cluster early-type galaxies because the convolution 
in age and metallicity of simple stellar populations with the chemical
enrichment tracks predicted by these models yield $U-V$ or $V-K$ colours 
which are significantly bluer than the observed values, $U-V\sim 1.6$, 
$V-K\sim 3.3$ for Coma or Virgo (Bower, Lucey \& Ellis 1992). 
The solution to this problem involves either 
pre-enrichment, i.e. shifting the zero-point of the metal 
abundance by assuming non-primordial infall (e.g. Sansom \& 
Proctor 1998), or imposing an initial stage 
of moderate star formation that enriches the interstellar medium, 
followed by a second stage where most of the stars (with 
non-primordial abundances) are formed. Several authors 
(Elbaz, Arnaud \& Vangioni-Flam 1995;
Chiappini, Matteucci \& Gratton 1997)
have suggested two-stage processes to solve this problem, 
as well as a top-heavy initial mass function 
for the first star forming stage that overproduces high mass 
stars for a prompt enrichment of the interstellar medium,
motivated by the fact that starbursts can be best explained by
top-heavy mass functions (Charlot et al. 1993).

In this paper we present a model which avoids these assumptions 
and solves the G-dwarf problem by considering infall of primordial 
gas with a rate given by a gaussian distribution. 
This infall combined with a linear 
Schmidt law results in a gaussian profile for the 
star formation rate which is a suitable way of parametrizing a 
strongly peaked starburst. Furthermore, the extension of this 
profile to disk galaxies, and in particular to our own galaxy, 
accounts for the observed metallicity distribution of stars 
(Rocha-Pinto \& Maciel 1996, Wyse \& 
Gilmore 1995):
the first stars are formed in a stage with a low SFR, but their
contribution to the enrichment of the IGM, along with an increasing
SFR with time yields a large population of stars with metallicities
close to solar. Once the IGM has reached a very large metallicity ---
after the peak in the SFR --- a low formation rate keeps the tally
of stars with a high metal content as low as observed. In this
scenario, the difference between early-type and late-type galaxies
would amount to an extra component in the SFR for late-type
systems, which continue to form stars at a slow rate 
(Ferreras \& Silk, in preparation). Hence, the work in 
this paper uses the same chemical enrichment equations (described 
in FS00) based on the formalism described by Tinsley (1980), 
with just four parameters describing the star formation process, 
namely:

\begin{itemize}
\item[$\bullet$] Star Formation Efficiency ($\ceff$): The 
  star formation rate is modelled by a linear Schmidt law, whose 
  proportionality constant determines the formation efficiency. 
  This parameter is a phenomenological approximation to the complex
  physics underlying star formation. Its inverse represents the 
  timescale in Gyr over which the SFR is extended for a sharply
  peaked infall of gas. In the Instantaneous Recycling 
  Approximation (IRA), the SFR ($\psi(t)$) and the infall 
  rate ($f(t)$) are related by:
\begin{equation}
\psi(t) = \ceff\int_0^\infty ds f(t-s)e^{-s/\tau_g},
\end{equation}
\begin{equation}
\tau_g = \frac{1}{\ceff\left[1-(1-\bout)R\right]}; 
\end{equation}
   $B_{\rm out}$ is the 
   ejected fraction of gas in outflows --- defined below --- and 
   $R$ is the returned fraction. Low values of $\ceff$ extend the 
   process of star formation, generating a larger age spread of the 
   stellar populations.
\item[$\bullet$] Gas Outflows ($\bout$): Part of the 
   metal-enriched gas in the IGM is heated by supernovae and ejected 
   from the galaxy, decreasing the effective yield. This will be 
   modulated by the gravitational potential well of the galaxy, so 
   that bright and massive galaxies have lower ejected fractions 
   and so, higher metal abundances (Larson 1974, Arimoto 
   \& Yoshii 1987). $\bout$ represents the fraction of gas being 
   ejected at each timestep.
\item[$\bullet$] Formation redshift ($z_F$): This parameter 
   describes the epoch --- $t(z_F)$ --- at which the infall 
   rate (or roughly the SFR) was maximum.
\item[$\bullet$] Infall timescale ($\tau_f$): Determines the 
   duration of infall, so that the infall rate can be written:
\begin{equation}
f(t,\tau_f,z_F) = \frac{1}{\tau_f\sqrt{2\pi}}\exp\left[
-\frac{\left(t-t(z_F)\right)^2}{2\tau_f^2}\right]
\end{equation}
\end{itemize}

For a given set of these four parameters ($\ceff$,$\bout$,
$z_F$,$\tau_f$) the equations give a chemical enrichment track 
which is used to convolve the simple stellar populations from 
the models of Bruzual \& Charlot (2000) both in 
age and metallicity, to obtain a spectral energy distribution 
for a given redshift. Motivated by recent results of the 
angular power spectrum of the Cosmic Microwave Background
(Melchiorri et al. 1999), we assume a flat cosmology with a 
cosmological constant ($\Omega_m=0.3$, $\Omega_\Lambda=0.7$, 
H$_0=60$ km s$^{-1}$Mpc$^{-1}$), although figure~1 shows the 
result for an open cosmology ($\Omega_\Lambda =0$) in the range 
of infall parameters ($\tau_f$ and $z_F$). The fiducial Initial 
Mass Function (IMF) chosen is a hybrid one between the Scalo (1986) 
and the Salpeter (1955) IMF with mass cutoffs $0.1 < M/M_\odot < 60$.
We have adopted the behaviour of the 
Scalo IMF for low stellar masses ($M<2M_\odot$) and a Salpeter 
IMF for the high mass end. The shallow slope of the former for 
low masses better explains the observations, whereas the 
steeper slope at high masses of the Salpeter IMF seem to explain 
better the stellar populations in starbursts. For comparison 
purposes we show the difference between a Salpeter IMF and our 
mass function when calculating the correlation between
mass-to-light ratio and mass (\S 5). 

Broadband colours are computed from the spectral energy distribution 
obtained for a given star formation history described by these four 
parameters ($\ceff$,$\bout$,$z_F$,$\tau_f$). The colour-magnitude
relation observed in local clusters is used as a constraint, so that
a one-to-one mapping is assumed between colour and absolute luminosity.
We have used the observations of Coma early-type galaxies
(ellipticals and lenticulars) by Bower et al. (1992), whose 
colour-magnitude relation has a scatter ($\pm 0.05$ mag) as small 
as the uncertainties expected from population synthesis models 
(Charlot, Worthey \& Bressan 1996). 
Hence, for a given $U-V$ colour obtained from the spectral energy 
distribution we infer an absolute luminosity:
\begin{equation}
   M_V = -\frac{(U-V) + 0.3830}{0.0871}.
\end{equation}
This correlation was obtained by a robust two-stage linear 
fit technique which computes a first estimate of slope and
zero point using absolute deviations (e.g. Press et al. 1992)
and then applies a least-squares fit to the resulting sample 
after culling points that deviate more than a given threshold.
This method prevents outliers from contributing significantly
to the final slope and zero point. In this case the number of
outliers was 4 out of 36 galaxies. Once the $V$-band absolute
luminosity is found, the total stellar mass is obtained using 
the predicted mass-to-light ratio.

%%%%%%%%%%%%%%%%%%%%%%%%%%%%%%%%%%%%%%%%%%%%%%%%%%%%%%%%%%%%%%%%%%%

\begin{figure}
 \epsfxsize=3.5in
 \epsffile{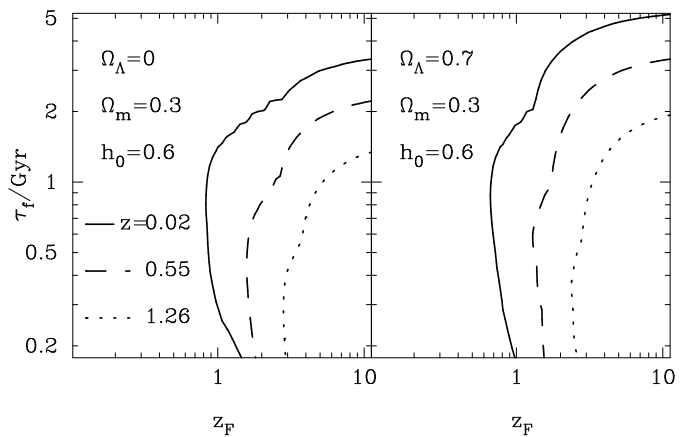}
 \caption{Allowed values of infall timescale ($\tau_f$) and formation
redshift ($z_F$) taking into account colour constraints for the brightest
galaxies in local and distant clusters: The $z=0.02$ 
constraint requires $U-V$ and $V-K$ to be correlated with a 
difference of just $\pm 0.05$ mag as observed in the Coma
cluster (Bower et al. 1992); the $z=0.55$ contour requires the 
F555W$-$F814W colour to be redder than 2.4 (Cl0016+16, 
Ellis et al. 1997), the $z=1.26$ contour imposes a colour 
$R-K >5.7$ as observed by Rosati et al. (1999) in cluster 
RX J0848.9+4452. The cosmology used assumes a matter
density $\Omega_m=0.3$ and $H_0=60$ km s$^{-1}$ Mpc $^{-1}$. 
The left and right panels show the contours for an 
open ($\Omega_\Lambda=0$) and flat ($\Omega_\Lambda=0.7$) 
cosmology, respectively.}
\label{f1}
\end{figure}

\section{Constraints from bright cluster galaxies}
The colour of the brightest galaxies in clusters at 
several redshifts is one of the strongest and most robust 
constraint that can
be imposed on the model parameters. Rest frame $U-V$ colour
is very sensitive to age (as well as metallicity). The red
$U-V$ colours of the brightest early-type systems observed 
in Coma ($U-V\sim 1.6$)
need a dominant population of old stars with a metallicity
around solar or higher. Hence, large infall timescales ($\tau_f$)
or low formation redshifts are readily ruled out since they
predict bluer $U-V$ colours. Furthermore, observations of 
clusters at moderate redshifts (e.g. Ellis et al. 1997;
Stanford, Eisenhardt \& Dickinson 1998; 
Dressler et al. 1999) 
have found a colour-magnitude relation consistent with passive 
evolution, imposing again an old age and high metal content. 
This motivates the need to use a low ejected fraction
($\bout \sim 0$) as well as a high star formation efficiency 
($\ceff\ga 5$) in order to reproduce the spectrophotometric 
properties of these galaxies. Figure~1 shows the constraint on 
infall parameters for the brightest galaxies using a grid 
of models with $\bout=0$ and
$\ceff=10$. The $z\sim 0$ constraint imposes a colour$-$colour correlation
between $U-V$ and $V-K$ as observed in Coma and Virgo (Bower
et al. 1992) with a scatter of $\pm 0.05$ mag. Notice that as
long as the bursting stage is short enough ($\tau_f\la 1$ Gyr),
formation redshifts as low as $z_F\sim 1$ can be accomodated
with the photometric data (Ferreras et al. 1999). In fact,
a simple stellar population (which is the extreme case of the
model described in this paper as $\tau_f\rightarrow 0$) gives
a $z=0$ color of $U-V\sim 1.5$ for a stellar population with solar
metallicity and formation redshift $z_F=1$. This corresponds
to a bright ($M_V\sim -21.6$) cluster galaxy.
The $z\sim 0.5$ constraint imposes a colour F555W$-$F814W
(which maps into restframe $U-V$) redder than $2.4$  for the
brightest cluster galaxies, as observed in Cl0016+16 ($z=0.55$, 
Ellis et al. 1997 ). The high redshift constraint
requires $R-K\ga 5.7$ as observed in cluster RX J0848.9+4452 
($z=1.26$, Rosati et al. 1999). 
These constraints rule out models with infall timescales 
$\tau_f\ga 1.5$ Gyr, which reduces the main star formation 
episode to less than $\sim 3$ Gyr. The formation redshift is 
less restricted, allowing the possibility of $z_F\ga 3$, or even 
$z_F\ga 2$ with a flat cosmology with a cosmological constant
($\Omega_\Lambda = 0.7$).

The constraint on the allowed volume of parameter space with 
respect to infall and its subsequent process of star formation 
shown in figure~1 is rather robust as it takes into account 
an observable with a very small uncertainty, namely the broadband 
colours of the most luminous galaxies.
Any significant process of star formation at later stages will
readily translate into bluer colours which are ruled out by
observations. The modelling of star formation rate as a gaussian
profile allows the extension of single burst models to processes
where several bursts of star formation take place, as in a
``gaseous'' merger scenario. The spectrophotometric predictions 
of a multi-burst scenario can be mimicked by a single gaussian 
infall rate with a longer time scale (Ferreras \& Silk 2000b).
Hence, if we consider RX J0848.9+4452 as a representative cluster
which will evolve into something like Coma or Virgo, then  
any hierarchical merging scenario should yield gaseous mergers
only at epochs earlier than $z_F\sim 3$. Merging stages at lower redshifts
should not trigger any significant star formation, for example
involving gas-poor progenitors and hot gas, as hinted at by 
observations of red merging galaxies in cluster 
MS$1054-03$ ($z=0.83$ Van Dokkum et al. 1999).

\begin{figure}
 \epsfxsize=3.5in
 \epsffile{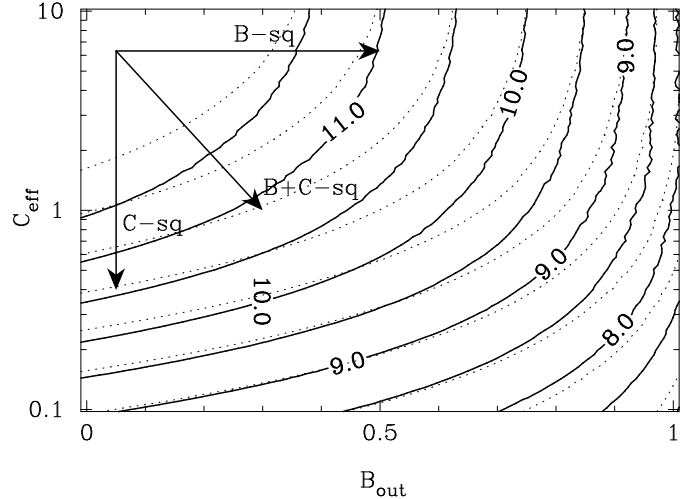}
 \caption{Mass contours --- labelled in terms of $\log M/M_\odot$ 
in a 2D parameter space spanned by $\bout$ and $\ceff$,
according to a model with $z_F=5$ and $\tau_f=1$ Gyr (solid lines)
or $\tau_f=0.5$ Gyr (dotted lines). The only region of 
parameter space allowed for the most massive galaxies requires a high star 
formation efficiency ($\ceff$) as well as a low ejected 
fraction ($\bout$). Also shown in the figure are the trajectories 
of a `'$\bout$ sequence'', a ``$\ceff$ sequence'' as well as 
a mixed ``$\bout +\ceff$ sequence'' as defined in the text.
Notice the mass contours do not differ much between models with infall
timescales of $\tau_f=0.5$ and $1$ Gyr for high SF efficiencies (top
part of the figure), as expected, since high values of $\ceff$ minimize 
the age spread of the stellar populations.}
 \label{f2}
\end{figure}

It is worth emphasizing at this point that the brightest galaxies
we are dealing with in this model represent the brightest 
``non-peculiar'' systems. Hence, the very brightest cluster galaxies
(BCGs) such as cDs should be excluded. In a recent paper, 
Arag\'on-Salamanca, Baugh \& Kauffmann (1998) analysed the 
evolution with redshift of a sample of BCGs in 25 clusters in the 
redshift range $0<z<1$. They found no evolution, or even a {\sl fading} 
of the absolute $K$-band luminosity which can be explained by a
significant change (a factor of 2) in the stellar mass of these galaxies
between redshifts $z=1$ and $0$. 
In the formalism of our model, the reddest --- thereby brightest ---
galaxies are obtained in the corner of parameter space associated with
high star formation efficiency ($\ceff\sim 5-10$). 
This implies the star formation rate will be approximately given by the 
infall rate $\psi(t)\sim f(t)$. For the allowed values of infall parameters
($\tau_f$, $z_F$) shown in figure~1, the change in stellar mass between
redshifts $z=1$ and $0$ is very small, and so this model would not
be compatible with BCGs. This type of galaxy requires a strong merger
rate at late times ($z\sim 1$). Furthermore, the mergers should undergo 
no star formation, involving progenitors with either very little gas or
hot gas. On the other hand, the evolution of the bulk of cluster
early-type systems was explored in a sample of 38 clusters in a similar
redshift range ($0.1<z<1$) by analysing the near-infrared luminosity 
function (De~Propris et al. 1999). A significant positive luminosity 
evolution was found for the Schechter parameter $K_*$ which traces 
the luminosity of the brightest galaxies. The evolution of $K_*$ is
found to be compatible with the passive evolution of a simple stellar
population, which is consistent with the models presented in this paper
for the range of infall parameters shown in figure~1. 

\section{Exploring parameter space}
Once we have fixed a region in parameter space for the brightest 
``non-peculiar'' cluster galaxies, we can extend the model
to all early-type systems by exploring the parameters describing
star formation efficiency and fraction of ejected gas in outflows.
Figure~2 shows a contour of stellar masses for two
realizations of infall  ($z_F=5$, $\tau_f=0.5$ Gyr, dotted lines)
and ($z_F=5$, $\tau_f=1$ Gyr, solid lines). The only constraint
imposed is the colour-magnitude relation in Coma in order to 
relate colour --- obtained from the star formation history described
by a chosen set of parameters ($\bout$, $\ceff$, $\tau_f$, $z_F$) 
--- and luminosity, or stellar mass. The contours
change slightly with infall parameters but the qualitative
behaviour is unchanged. We need {\sl both} a high star formation
efficiency and a low ejected fraction in outflows to account
for the most massive galaxies. Lowering the SF efficiency will
generate a significant spread in stellar ages that will yield
colours that are too blue compared with the observations. On the 
other hand, a higher ejected fraction will decrease the effective 
yield, so that the average metallicity will be lower, blueing the
predicted colours with respect to the observations. Hence,
regardless of the choice of parameters, we find the brightest
cluster galaxies to lie in a well-defined corner of the
4-dimensional parameter space explored in this paper. Lower mass
galaxies are harder to identify because of a degeneracy between
outflow rates and star formation efficiency. The blueing of galaxies
as we progress down in luminosity along the colour-magnitude
relation can be explained either by decreasing the average metallicity
of the stellar populations (i.e. a range of ejected fractions, $\bout$) 
or by increasing the fraction of younger stars (i.e. a range of
SF efficiencies, $\ceff$). The real picture will likely involve a 
range both in $\bout$ and $\ceff$. At low redshift it is
hard to disentangle the contribution from these two parameters,
mainly caused by the age-metallicity degeneracy 
(Worthey 1994). However, these two alternative 
mechanisms will predict different evolutions with lookback time: 
as we go to higher redshift, the sequence driven by SF efficiency
(hereafter $\ceff$ sequence) will show stronger evolutionary scars 
because of the larger scatter in age, whereas the sequence driven 
by outflows (hereafter $\bout$ sequence) will be unaffected until the 
observations go back to the epoch of star formation.

\begin{figure}
 \epsfxsize=3.5in
 \epsffile{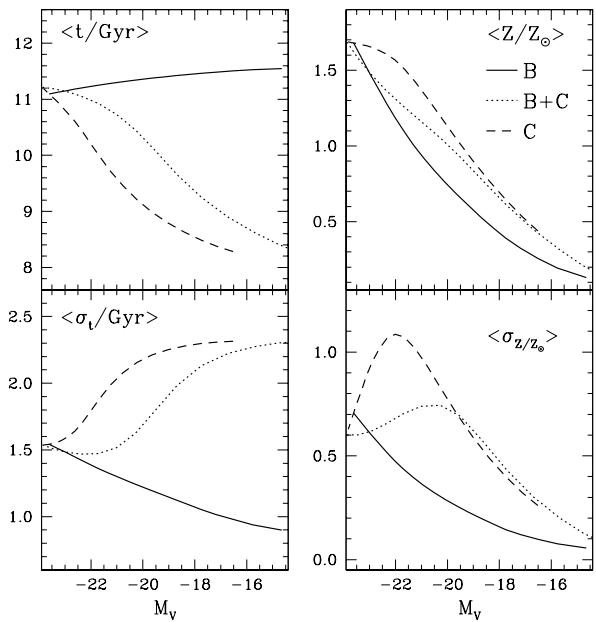}
 \caption{Distribution of stellar ages and metallicities according
to the three sequences defined in the text. The top panels show the
average age ({\sl left}), and metallicity ({\sl right}), whereas
the bottom panels show the standard deviation of the distributions.
Notice a ``$\bout$ sequence'' generates a pure metallicity sequence
for cluster ellipticals, with all galaxies being roughly coeval,
in agreement with the observations in the Fornax cluster
(Kuntscher \& Davies 1998) The closest model to a pure age 
sequence is a ``$\ceff$ sequence'', where a range of star formation 
efficiencies result in a significant average age spread with 
respect to luminosity. However, there is also an important 
range of metallicities. A pure age sequence is ruled
out because of chemical enrichment.}
 \label{f3}
\end{figure}

Figure~3 shows the predicted distributions of stellar ages 
({\sl left}) and metallicities ({\sl right}). The average
({\sl top}) and standard deviation ({\sl bottom}) of these
distributions are shown for three sequences: $\bout$,$\ceff$ 
and $\bout +\ceff$ driven by outflows, star formation efficiency 
and a mixture of both, respectively. All of these models
have $\tau_f=1$ Gyr, $z_F=5$; the $\bout$ sequence fixes
$\ceff =5$; the $\ceff$ sequence fixes $\bout =0$; and the
mixed $\bout +\ceff$ sequence imposes a correlation between 
these two parameters. We expect the star formation efficiency to be
higher in more massive galaxies (where the ejected fraction should
be lower), hence we have chosen: $\log\ceff = 1-3\cdot\bout$,
as an illustrative example. An outflow-driven sequence generates 
a pure metallicity sequence where the stellar populations of all 
galaxies are coeval. This is the model favored by the analysis
of the colour-magnitude relation by Kodama et al. (1998) for a 
sample of 17 clusters in a wide redshift range ($0.3<z<1.3$).
In a $\bout$ sequence the colour-magnitude 
relation is  due to a pure mass-metallicity correlation, so that 
the brightest galaxies have higher metal abundances. 
On the other hand, a $\ceff$ or a $\bout +\ceff$ model represents 
mixed age+metallicity sequences as it generates a significant 
age spread of the stellar populations as well as a metallicity range. 
A pure age sequence explored in other papers (e.g. Kodama \& 
Arimoto 1997) is ruled out when including chemical enrichment for
any reasonable assumption about the star formation history of the 
galaxy. The low metallicity expected in the fainter galaxies can
therefore be explained either by a lower effective yield caused
by gas outflows, or by a low star formation efficiency that slows
down the process of enrichment. Nagashima \& Gouda (1999) mention
the effect of the UV background radiation as a possible alternative to
supernovae winds to suppress chemical enrichment. 
However, in this scenario it seems hard to motivate the
correlation found with respect to luminosity (e.g. a stronger feedback
should be expected from the UV background for the fainter galaxies).
Figure~3 shows that a $\ceff$ sequence implies a $\sim 3$ Gyr age 
difference between the brightest systems and $M_V\sim -16$ galaxies with a larger 
age spread for the fainter galaxies, as expected for a system with 
a lower star formation efficiency. This model would thus be 
consistent with the smaller age and metallicity scatter found in
Coma ellipticals (J\o rgensen 1999). This age difference is negligible
for local clusters, since the predicted youngest stars are still
too old ($\sim 8$ Gyr) to be detected spectrophotometrically.
However, predictions between a coeval sequence and
a $\ceff$ sequence at lookback times $\sim 8$  Gyr (i.e. $z\sim 1$)
differ significantly as shown below (\S5). A $\bout$ sequence 
predicts a small age and metallicity spread in faint 
ellipticals. In this case the observed photometric scaling 
relations are just driven by the average metallicity.

%%%%%%%%%%%%%%%%%%%%%%%%%%%%%%%%%%%%%%%%%%%%%%%%%%%%%%%%%%%%%%%%%%

\section{Mass-to-Light ratios: Stellar versus Observed}

The evolution of the stellar mass-to-light ratio is a
suitable probe for the analysis of the ages of the
stellar populations in galaxies because of its 
weak dependence on metallicity (especially when measured in 
NIR passbands). However, observed $M/L$ ratios carry large
uncertainties and systematic offsets as they are computed 
from central velocity dispersions ($\sigma_0$), sizes ($r_e$)
and surface brightnesses ($\Sigma_e$), 
requiring a prescription for the structure in order to infer 
masses and luminosities. From a simple dimensional analysis, we
can write the mass and luminosity of a galaxy as:
\begin{equation}
M = \alpha \sigma_0^2 r_e
\end{equation}
\begin{equation}
L = \beta r_e^2 \Sigma_e
\end{equation}
The proportionality constant $\beta$ only depends on the way the surface
brightness is defined with respect to the effective size. However, 
$\alpha$ depends on the structure of the galaxy, which requires a
dynamical model (e.g. King 1966; Bender, Burstein \& Faber 1992), 
usually assumed to be invariant with regard to galaxy mass 
or size (i.e. homologous). Furthermore, a comparison between observations
and model predictions depends on a prescription that relates 
stellar and total matter. Despite all these caveats, mass-to-light
ratios are one of the best candidates for breaking the age-metallicity
degeneracy.

\begin{figure}
 \epsfxsize=3.5in
 \epsffile{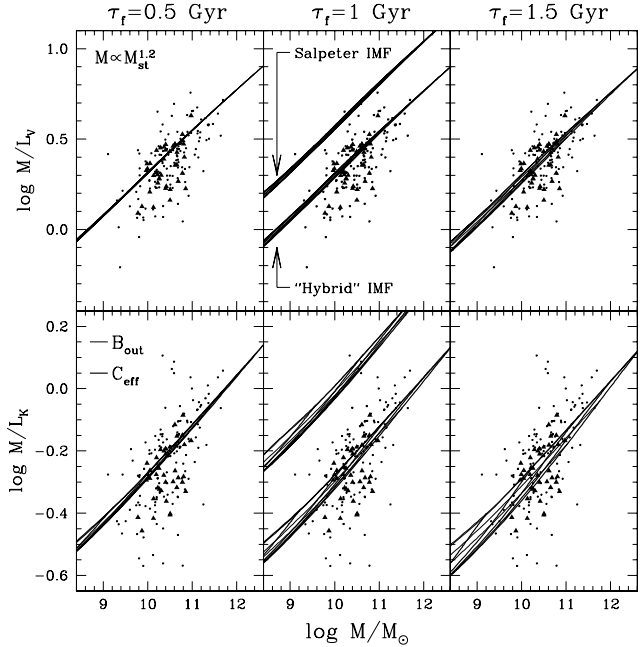}
 \caption{Model predictions of the correlation between stellar 
mass-to-light ratio and mass in two different passbands:
$V$ band ({\sl top}) and $K$ band ({\sl bottom}). Notice the NIR
observable is much more sensitive to age and thus changes 
significantly with infall timescale. The $\tau_f=1$ Gyr case also
displays the predictions for a Salpeter IMF, rather high in comparison
with observations. The triangles show Coma cluster ellipticals from
Mobasher et al. (1999), whereas the dots show ellipticals in a
sample of 11 clusters extracted from Pahre (1999).
The predictions --- which just give stellar masses and mass-to-light
ratios --- have been transformed to total mass and $M/L$ using a 
non-linear correlation between total mass (inferred by observations), 
and stellar mass (given by population synthesis models): 
$M_{\rm TOT}\propto M_{\rm St}^{1.2}$. The grid comprises a set
of $\bout$ sequences (thin) with 
fixed $\ceff=\{0.1,0.5,1,5,10\}$ and a set of $\ceff$ sequences 
(thick) with fixed $\bout=\{0,0.2,0.4,0.6,0.8\}$.}
 \label{f4}
\end{figure}

Figure~4 shows the predicted $M/L$ ratios in two passbands:
optical ($V$) and NIR ($K$) for a grid of $\bout$ (thin) and $\ceff$
(bold) sequences. A $\bout +\ceff$ mixed sequence would be represented by a
trajectory between the grid spanned by these lines. The
results are shown for three different infall timescales
$\tau_f=0.5,1,1.5$ Gyr. The solid triangles are the observations 
of Coma cluster galaxies from Mobasher et al. (1999), whereas
the dots represent ellipticals in 11 clusters from the sample of 
Pahre (1999). The slope of the correlation between 
mass-to-light ratios and masses is $0.24$ in the $V$ band and
$0.14$ in $K$ band, whereas purely stellar $M/L$ ratios yield slopes
around $0.1$ and $0.0$ in the $V$ and $K$ bands, respectively. 
This mismatch cannot be related to a different stellar population. 
Being a strongly age-dependent observable, the only way to generate
the observed slopes in $M/L$ would require a stellar age spread
with respect to galaxy mass whose restframe $U-V$ colour would
significantly disagree with the colours in moderate redshift
cluster ellipticals. As a simple check to confirm this, we
used simple stellar populations (SSP, i.e. no age or metallicity
spread). If we relate the brightest galaxies ($M_V\sim -22.5$) 
to a SSP with an age of 12 Gyr and a super-solar metallicity 
$Z=1.5Z_\odot$, the required age for the SSP for a 
$M_V=-19$ galaxy which would  explain the observed
slope $M/L_V \propto M^{0.24}$ should be $t=4$ Gyr ($Z=Z_\odot$)
or $t=5$ Gyr ($Z=Z_\odot/2$). These two possibilities
yield $U-V$ colours of $1.20$ and $1.07$ respectively, so that
the latter is barely consistent with the observed broadband
photometry in nearby clusters (Bower et al. 1992). However, this
hypothesis fails when considering moderate redshift clusters:
for instance, cluster Cl0016+16 ($z\sim 0.5$) has a well-defined
red-envelope in a large range of luminosities (Ellis et al. 1997), 
yet the lookback time ($\sim 6.2$ Gyr in our adopted cosmology) would 
be larger than the expected age for the stars in the fainter galaxies !

Hence, the only plausible way to solve
this slope mismatch is by imposing a non-linear correlation
between total mass (including dark matter) and stellar mass.
A simple power law $M\propto M_{\rm ST}^{1.2}$ (FS00) has been 
used in figure~4 to transform predicted stellar masses and 
mass-to-light ratios to total masses and $M/L$. The
agreement is good in both bands, which means the observed slope 
difference in the fundamental plane between passbands can be 
explained by the stellar populations alone. This result agrees with
the analysis of J\o rgensen (1999) who refers to a higher fraction 
of dark matter in the most massive ellipticals in order to account
for the mismatch between age estimates using $M/L$ ratios (which 
involves the total mass of the galaxies) or $H\beta_G$ indices 
(involving just the stellar component). However, one should bear 
in mind that there are alternative explanations for this slope 
mismatch, such as a systematic variation of the initial mass function
or the breaking of the homology for the structural parameters
of early-type systems (J\o rgensen 1999; Pahre et al. 1998). 
Graham \& Colless (1997) examined the effect of a non-homologous
light profile in 26 early-type systems in the Virgo cluster and 
found no differences in the fundamental plane from estimates 
using a universal de Vaucouleurs profile.

Notice the scatter of the correlation is not the minimum one can obtain
when computing the fundamental plane. The dependence of the fundamental
plane on the observables (velocity dispersion, surface brightness 
and galaxy size) is not precisely the one inferred for the correlation 
between mass-to-light ratio and mass. This scatter has an observational
component as well as an intrinsic contribution. The intrinsic
part can be related to differences in either the stellar ages 
or the initial mass function of galaxies with the same mass.
The former can be caused by slightly different formation redshifts
or infall timescales. Notice the predictions of $M/L$ ratios 
for three different infall timescales in figure~4. A variation of 
1 Gyr in infall timescale can account for changes in $\log M/L$
roughly of order $0.1$ dex. Furthermore, a non-universal IMF
will add a variation around $0.2-0.3$ dex. The top-center
panel displays the predictions for our fiducial ``hybrid'' IMF
and a Salpeter IMF with the same mass cutoffs ($0.1<M/M_\odot<60$).
A different cutoff or slope at the high mass end will not change
the result very much since these changes only have an effect on 
metallicity. However, changes in the low mass end of the IMF will
significantly alter the mass-to-light ratio. Our ``hybrid IMF'' 
behaves like a Scalo function at low masses, whereas a Salpeter
mass function produces a large fraction of low mass stars, thereby
increasing the stellar $M/L$ ratio.

\begin{figure}
 \epsfxsize=3.5in
 \epsffile{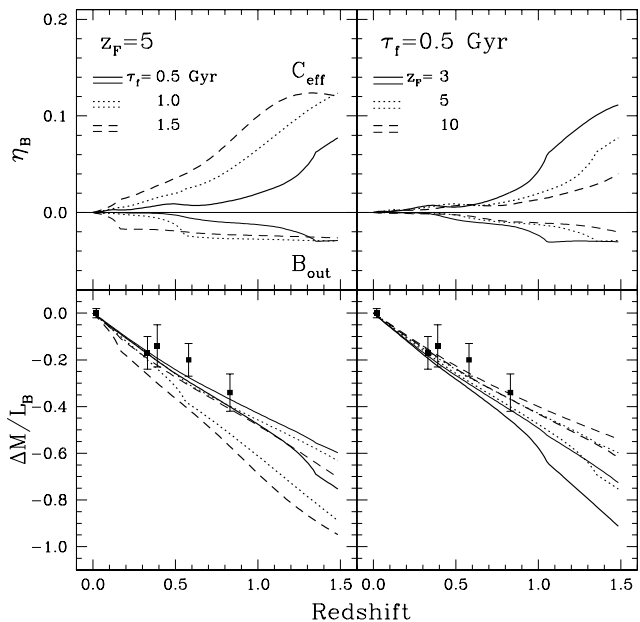}
 \caption{Predicted evolution with observed redshift of the slope 
({\sl top}) and zero point ({\sl bottom}) of the correlation between
mass-to-light ratio and mass in the $B$ band, in models driven 
by star formation efficiency ($\ceff$, thick lines) or outflows
($\bout$, thin lines). $\eta_B$ is defined
as the difference between the slope in $\log M/L_B$ vs $\log M$ 
at $z>0$ and $z=0$. The data points in the bottom panels show 
observed shifts in the zero point for clusters: Coma ($z=0.02$), 
Cl 1358+62 ($z=0.33$), Cl 0024+16 ($z=0.39$), MS 2053+03 ($z=0.58$) 
and MS1054-03 ($z=0.83$) from the compilation in Van Dokkum 
et al.(1998).}\label{f5}
\end{figure}

The different age distributions obtained for the $\bout$ and $\ceff$ 
sequences generate different predictions for the evolution
of the slope of the correlation between stellar $M/L$ ratio 
and mass. We have shown above that a $\bout$ sequence (driven by
outflows) is equivalent to a metallicity sequence. Hence, 
a very small slope change is expected since mass-to-light ratios
are not very sensitive to metal abundance. Furthermore, the
decrease in $M/L$ will be greater at higher metallicities (i.e.
for the brightest galaxies), so that the correlation will get
flatter. In order to quantify the slope change, an index $\eta_X$
is defined as the slope change between redshift $z>0$ and
$z=0$ using mass-to-light ratios in the $X$ passband, namely:
\begin{equation}
\eta_X(z) \equiv \left.\frac{\Delta\log M/L_X}{\Delta\log M}\right|_{z} -
\left.\frac{\Delta\log M/L_X}{\Delta\log M}\right|_{z=0}
\end{equation}
Hence, a $\bout$ sequence predicts small negative values for $\eta_X(z)$.
On the other hand, a $\ceff$ sequence is driven by age, so that the
faintest galaxies --- which will have a younger stellar population
--- will decrease their $M/L$ ratios with redshift faster than 
the brighter (and older) galaxies. This corresponds to a more
significant and positive value for $\eta_X(z)$. Figure~5 shows
the evolution of the slope ({\sl top panels}) and zero point
({\sl bottom panels}) of the correlation between mass-to-light
ratio and mass for a range of infall timescales ({\sl left})
or formation redshifts ({\sl right}). The points show the
observations of a few clusters in restframe $B$ band: 
Coma ($z=0.02$), Cl 1358+62 ($z=0.33$), 
Cl 0024+16 ($z=0.39$), MS 2053+03 ($z=0.58$) and 
MS1054-03 ($z=0.83$) from the compilation in Van Dokkum et al.
(1998). Any sequence ($\bout$, $\ceff$, $\bout +\ceff$) 
will predict a decrease of the zero point as the stellar 
populations get younger at higher redshift, although the zero point 
decreases faster for a $\ceff$ sequence, 
for which the average stellar population 
is younger. Current observations of moderate redshift cluster
ellipticals allow an estimate of the zero point but not of the
slope with enough accuracy. For instance, the study of 
J\o rgensen et al. (1999) with clusters at low and moderate
redshifts seem to indicate a steepening of the correlation
between $M/L$ and mass, thereby favouring a $\ceff$ sequence. 
However, selection effects could actually
mimic such behaviour. Forthcoming spectroscopic observations
of cluster ellipticals will enable us to estimate this slope,
thereby allowing us to determine the importance of outflows and
star formation efficiency in the evolution of cluster early-type
galaxies.

%%%%%%%%%%%%%%%%%%%%%%%%%%%%%%%%%%%%%%%%%%%%%%%%%%%%%%%%%%%%%%%%%%%%%

\section{Discussion}

A simple phenomenological treatment is described in this paper,
where the mechanisms underlying the evolution of galaxies
are reduced to a set of four parameters. The star formation
rate is assumed to follow a linear Schmidt law whose
proportionality constant is used to describe a varying star 
formation efficiency ($\ceff$). The supply of primordial gas
fuelling star formation is controlled by gaussian infall 
characterised by the epoch at which the infall rate is 
maximum ($z_F$), and the width of the gaussian profile
gives a characteristic infall timescale ($\tau_f$). The 
model is allowed to eject a fraction 
($\bout$) of the enriched gas, thereby lowering the 
effective yield. A first stage in this analysis involves
finding a suitable pair of infall parameters 
($z_F$,$\tau_f$) which are capable of generating 
the restframe $U-V$ colours of the brightest cluster 
galaxies. The colour constraints imposed by the reddest 
(and brightest) galaxies 
in clusters at moderate and high redshift allow us to discard long
infall timescales and recent star formation epochs. One could
argue that the constraint on the star formation history
of the brightest systems
need not be the same as for less massive ellipticals.
However, the current most plausible scenarios for galaxy formation
assume either a simultaneous process of star formation regardless
of galaxy mass, or a hierarchical structure where the most massive
galaxies might have undergone the {\sl latest} bursts of star formation.
Furthermore, the possibility of an ``inverted-hierarchical'' scenario 
can still be accomodated in this model, as long as the constraints
imposed ($z_F\ga 3$,$\tau_f\ga 1$ Gyr) are held even for the
less massive galaxies. 

Out of the four parameters considered in the model, we
have found that --- within the framework of this model --- 
only the SF efficiency ($\ceff$) and the ejected fraction 
in outflows ($\bout$) help determine the mass sequence in cluster 
early-type galaxies. Infall parameters ($z_F$,$\tau_f$) do not 
appreciably change much the output unless very recent stages of 
star formation are included. However, this will result in 
restframe $U-V$ colours that are in contradiction with observations. 
The efficiency parameter generates a significant spread in the
age distribution of stars, although mixed with a metallicity
range (any model with a reasonable IMF must include this
range of abundances). Alternatively, a range of outflow 
rates result in a range of metallicities with no significant 
spread in ages. Both mechanisms are degenerate in local clusters
because the average age of the stellar populations predicted for 
any model at $z=0$ are too old to be able to disentangle the effects 
of age and metallicity. However, the predictions of age-sensitive 
observables at high redshift differ noticeably for these two sequences. 
Unfortunately, present data is not capable of ruling out 
one model against the other, but the continuing flow of data 
from clusters at moderate-to-high redshift will eventually 
enable us to break this degeneracy. A sequence driven by efficiency 
($\ceff$ sequence) predicts a steepening of the slope of
the fundamental plane, whereas a $\bout$ sequence --- driven by
outflows --- predicts no change or a slight decrease of 
this slope. High precision observations of the dynamical
and spectrophotometric properties of cluster galaxies
at high redshift will confirm the importance of either age or
metallicity in the mass range of cluster ellipticals.

\section*{Acknowledgments}

The authors would like to thank the anonymous referee for useful
comments and suggestions.

\bsp

\label{lastpage}

\end{document}